\documentclass[twoside,twocolumn,9pt]{article}
\usepackage{extsizes}
\usepackage[super,sort&compress,comma]{natbib} 
\usepackage[version=3]{mhchem}
\usepackage[left=1.5cm, right=1.5cm, top=1.785cm, bottom=2.0cm]{geometry}
\usepackage{balance}
\usepackage{mathptmx}
\usepackage{sectsty}
\usepackage{graphicx} 
\usepackage{lastpage}
\usepackage[format=plain,justification=justified,singlelinecheck=false,font={stretch=1.125,small,sf},labelfont=bf,labelsep=space]{caption}
\usepackage{float}
\usepackage{fancyhdr}
\usepackage{fnpos}
\usepackage{epstopdf}
\usepackage{amsmath}
\usepackage{amssymb}
\usepackage{mathtools}
\usepackage{etoolbox}
\usepackage{tikz}

\usepackage[english]{babel}
\addto{\captionsenglish}{%
  
}
\usepackage{array}
\usepackage{droidsans}
\usepackage{charter}
\usepackage[T1]{fontenc}
\usepackage{setspace}
\usepackage[compact]{titlesec}
\usepackage{hyperref}

\usepackage{sidecap}% to put the caption on the side
\usepackage{placeins}% to force figure position

\usepackage{epstopdf}%This line makes .eps figures into .pdf - please comment out if not required.

\definecolor{cream}{RGB}{222,217,201}

\hypersetup{colorlinks=true, citecolor=blue, urlcolor=blue, linkcolor=blue}
\newrobustcmd*{\mycircle}[1]{\tikz{\filldraw[draw=#1,fill=#1] (0,0) circle [radius=0.08cm];}}
\newrobustcmd*{\myholowcircle}[1]{\tikz{\filldraw[draw=#1,fill=white] (0,0) circle [radius=0.07cm];}}
\newrobustcmd*{\myholowsquare}[1]{\tikz{\filldraw[draw=#1,fill=white] (0,0) rectangle ++(4pt,4pt);}}

\begin{document}

\pagestyle{fancy}
\thispagestyle{plain}
\fancypagestyle{plain}{
%%%HEADER%%%
\renewcommand{\headrulewidth}{0pt}
}
%%%END OF HEADER%%%

%%%PAGE SETUP - Please do not change any commands within this section%%%
\makeFNbottom
\makeatletter
\renewcommand\LARGE{\@setfontsize\LARGE{15pt}{17}}
\renewcommand\Large{\@setfontsize\Large{12pt}{14}}
\renewcommand\large{\@setfontsize\large{10pt}{12}}
\renewcommand\footnotesize{\@setfontsize\footnotesize{7pt}{10}}
\makeatother

\renewcommand{\thefootnote}{\fnsymbol{footnote}}
\renewcommand\footnoterule{\vspace*{1pt}% 
\color{cream}\hrule width 3.5in height 0.4pt \color{black}\vspace*{5pt}} 
\setcounter{secnumdepth}{5}

\makeatletter 
\renewcommand\@biblabel[1]{#1}            
\renewcommand\@makefntext[1]% 
{\noindent\makebox[0pt][r]{\@thefnmark\,}#1}
\makeatother 
\renewcommand{\figurename}{\small{Fig.}~}
\sectionfont{\sffamily\Large}
\subsectionfont{\normalsize}
\subsubsectionfont{\bf}
\setstretch{1.125} %In particular, please do not alter this line.
\setlength{\skip\footins}{0.8cm}
\setlength{\footnotesep}{0.25cm}
\setlength{\jot}{10pt}
\titlespacing*{\section}{0pt}{4pt}{4pt}
\titlespacing*{\subsection}{0pt}{15pt}{1pt}
%%%END OF PAGE SETUP%%%

%%%FOOTER%%%
\fancyfoot{}
\fancyfoot[LO,RE]{\vspace{-7.1pt}\includegraphics[height=9pt]{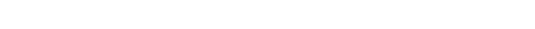}}
\fancyfoot[CO]{\vspace{-7.1pt}\hspace{11.9cm}\includegraphics{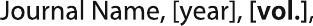}}
\fancyfoot[CE]{\vspace{-7.2pt}\hspace{-13.2cm}\includegraphics{head_foot/RF}}
\fancyfoot[RO]{\footnotesize{\sffamily{1--\pageref{LastPage} ~\textbar  \hspace{2pt}\thepage}}}
\fancyfoot[LE]{\footnotesize{\sffamily{\thepage~\textbar\hspace{4.65cm} 1--\pageref{LastPage}}}}
\fancyhead{}
\renewcommand{\headrulewidth}{0pt} 
\renewcommand{\footrulewidth}{0pt}
\setlength{\arrayrulewidth}{1pt}
\setlength{\columnsep}{6.5mm}
\setlength\bibsep{1pt}
%%%END OF FOOTER%%%

%%%FIGURE SETUP - please do not change any commands within this section%%%
\makeatletter 
\newlength{\figrulesep} 
\setlength{\figrulesep}{0.5\textfloatsep} 

\newcommand{\topfigrule}{\vspace*{-1pt}% 
\noindent{\color{cream}\rule[-\figrulesep]{\columnwidth}{1.5pt}} }

\newcommand{\botfigrule}{\vspace*{-2pt}% 
\noindent{\color{cream}\rule[\figrulesep]{\columnwidth}{1.5pt}} }

\newcommand{\dblfigrule}{\vspace*{-1pt}% 
\noindent{\color{cream}\rule[-\figrulesep]{\textwidth}{1.5pt}} }

\makeatother
%%%END OF FIGURE SETUP%%%

%%%TITLE, AUTHORS AND ABSTRACT%%%
\twocolumn[
  \begin{@twocolumnfalse}
{\includegraphics[height=30pt]{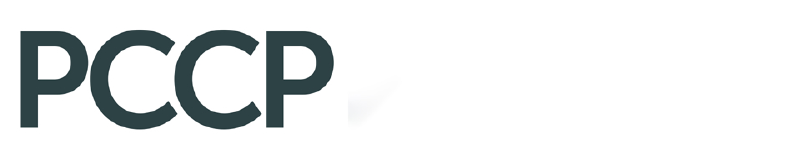}\hfill\raisebox{0pt}[0pt][0pt]{\includegraphics[height=55pt]{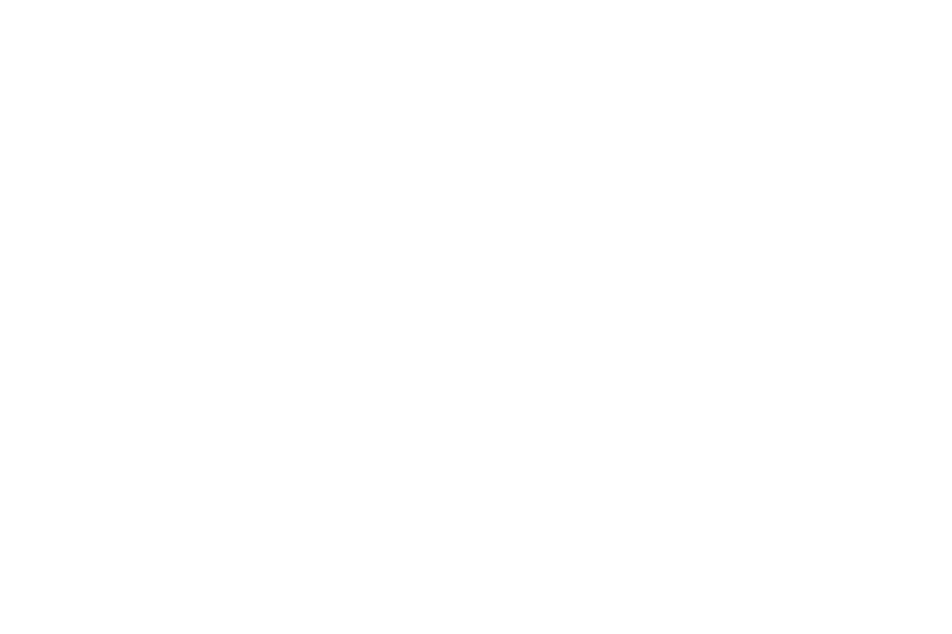}}\\[1ex]
\includegraphics[width=18.5cm]{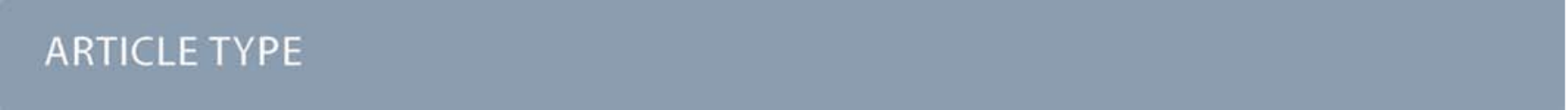}}\par
\vspace{1em}
\sffamily
\begin{tabular}{m{4.5cm} p{13.5cm} }
%\orcidlink{https://orcid.org}

\includegraphics{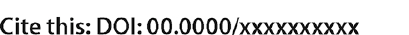} & \noindent\LARGE{\textbf{Temperature dependence in fast-atom diffraction at surfaces}} \\%Article title goes here instead of the text "This is the title"
\vspace{0.3cm} & \vspace{0.3cm} \\

 & \noindent\large{Peng Pan$^{a}$, Maxime Debiossac$^{a}$  and Philippe Roncin$^{a}$ }\\%Author names go here instead of "Full name", etc.

\includegraphics{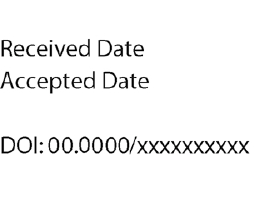} & \noindent\normalsize{Grazing incidence fast atom diffraction at crystal surfaces (GIFAD or FAD) has demonstrated coherent diffraction both at effective energies close to one eV ($\lambda_\perp\approx$ 14 pm for He) and at elevated surface temperatures offering high topological resolution and real time monitoring of growth processes. This is explained by a favorable Debye-Waller factor specific to the multiple collision regime of grazing incidence. This paper presents the first extensive evaluation of the temperature behavior between 177 and 1017 K on a LiF surface. Similarly to diffraction at thermal energies, an exponential attenuation of the elastic intensity is observed but the maximum coherence is hardly limited by the attraction forces. It is more influenced by the surface stiffness and appears very sensitive to surface defects.} \\%The abstrast goes here instead of the text "The abstract should be..."

%\item \href{https://orcid.org/0000-0002-7407-9474}{\textcolor{orcidlogocol}{\aiOrcid} \hspace{2mm} orcid.org/0000-0002-7407-9474}

\end{tabular}

 \end{@twocolumnfalse} \vspace{0.6cm}

  ]
%%%END OF TITLE, AUTHORS AND ABSTRACT%%%

%%%FONT SETUP - please do not change any commands within this section
\renewcommand*\rmdefault{bch}\normalfont\upshape
\rmfamily
\section*{}
\vspace{-1cm}

%%%FOOTNOTES%%%

\footnotetext{\textit{$^{a}$~Address, Universit\'{e} Paris-Saclay, CNRS, Institut des Sciences Mol\'{e}culaires d'Orsay (ISMO), 91405 Orsay, France}}
%\footnotetext{\textit{$^{b}$~Address, Address, Town, Country. }}

%\footnotetext{\ddag~Additional footnotes to the title and authors can be included \textit{e.g.}\ `Present address:' or `These authors contributed equally to this work' as above using the symbols: \ddag, \textsection, and \P. Please place the appropriate symbol next to the author's name and include a \texttt{\textbackslash footnotetext} entry in the the correct place in the list.}

%%%END OF FOOTNOTES%%%

%%%MAIN TEXT%%%%

\section{Introduction}
\begin{figure}[ht]
	\includegraphics[width=1.0\linewidth,trim={0 0mm 0 1mm},clip,draft = false]{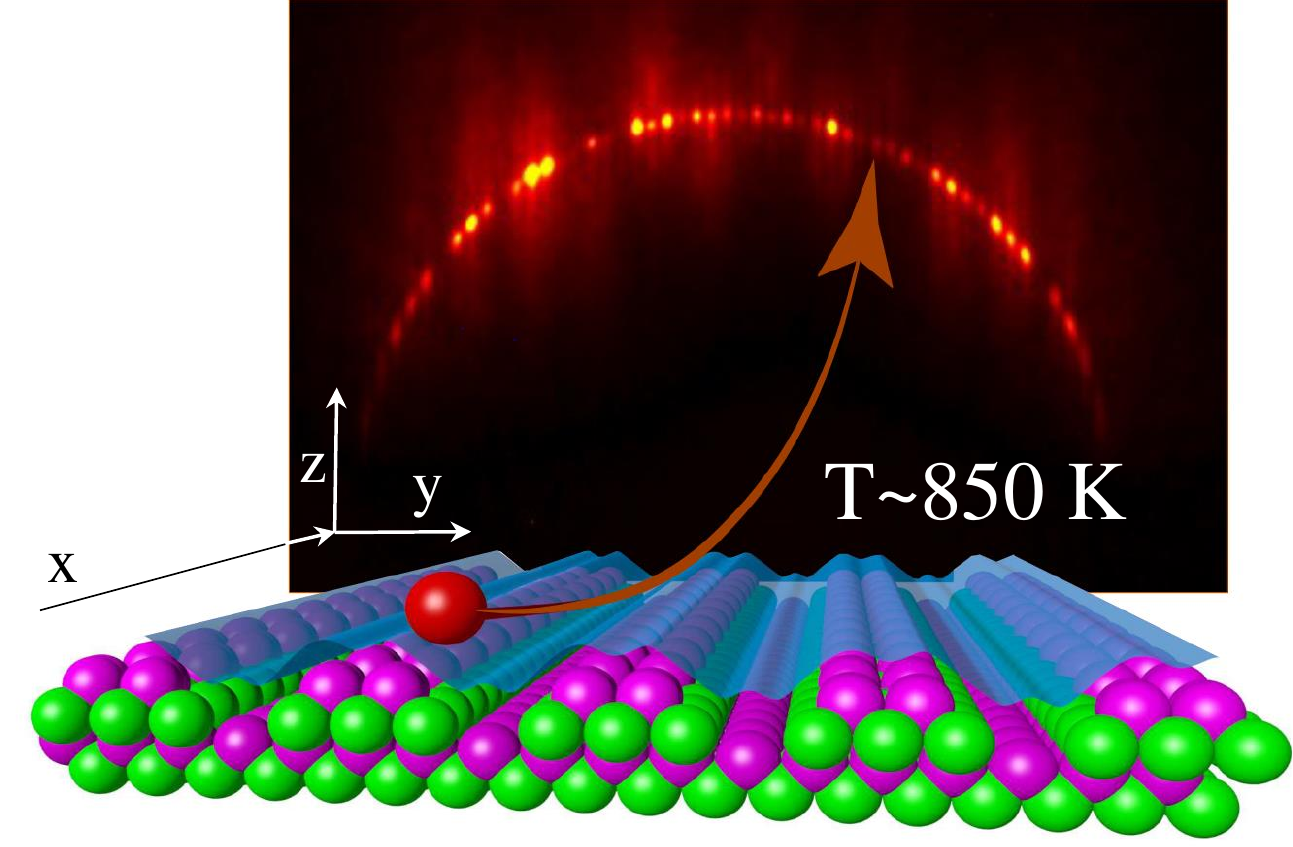}
	\caption{Schematic view of a GIFAD setup, the well-aligned row of atoms acts as a grating for the projectile atomic wave. A detector located almost a meter downstream records the diffraction image. The one here, taken from Ref.\cite{Debiossac_PRB_2014} was recorded directly in a MBE vessel on a GaAs(001) surface at $\sim$ 850 K. The bright spots corresponding to elastic diffraction are located on the Laue circle of energy conservation. \label{fgr:fig1}}
\end{figure}

The characterization of new materials requires a variety of techniques to analyze their physical and chemical properties. 
%In several cases such as 2D materials, surface sensitive techniques are needed to extract the behavior of the topmost layer.
These can be real space microscopic techniques such as scanning tunneling microscopy and atomic force microscopy but also reciprocal space techniques using X-rays, neutrons, electrons, or atoms.
Since atoms with kinetic energies below a few eV cannot penetrate below the surface, thermal energies atom scattering (TEAS), also known as helium atom scattering (HAS), is a valuable tool to investigate surfaces and 2D materials \cite{Farias_2019,Tamtogl_2021}. 
It is insensitive to the presence of magnetic or electric fields and does not induce any direct damage or charging of the surface.
However, its geometry is not compatible with a standard molecular beam epitaxy (MBE) vessel which requires that no instrument prevents the gas from the evaporation cells to reach the surface.
MBE also requires elevated surface temperatures in order to control the mobility of the deposited atom or molecule so that these can reach an optimum location in a reasonable timescale without being trapped too long in undesirable sites \cite{Dreiser_PRB2021}.
In this context, grazing incidence fast atom diffraction (GIFAD\cite{Rousseau_2007} or FAD\cite{Schuller_2007}) which is the high energy, grazing incidence counterpart of TEAS offers decisive advantages; it has shown to be compatible with a harsh UHV environment and with the MBE geometry. At variance with TEAS, the full diffraction image can be recorded in seconds so that it can be used as an real time diagnostic of the structure of the terminal layer. 
More unexpected, diffraction can be recorded on a surface at elevated temperatures.
This is illustrated in Fig.\ref{fgr:fig1} taken from Ref \cite{Debiossac_PRB_2014} and recorded inside a MBE vessel with a GaAs surface around 850 K, whereas TEAS is mainly performed at low temperatures. 
%\textcolor{red}{It has been reported the morphology strongly depends on the growth temperature, for LiF/Ag(100) model system, when the deposition occurs in high temperature(500K), LiF forms compact square islands\cite{Dreiser_PRB2021}}.
This interesting aspect of GIFAD to operate at elevated surface temperatures is poorly documented \cite{Rousseau_2008,Frisco_2019,Frisco_2020} and no systematic experimental investigation has been reported.

This paper presents experimental investigations on temperature dependence under a wide variety of experimental conditions of energy, angle of incidence and temperatures for helium atoms impinging on a LiF surface.
The paper is organized as follows, the experimental setup is described in Sec.\ref{ch:setup} with the protocol used to transform raw data into transverse and polar scattering profiles from which the coherence ratio is defined.
%Various images and profiles recorded at different temperatures are shown and described.
In Sec.\ref{ch:phi_scan}, the strategy adopted to performed stable temperature variations is presented before an introduction to theoretical aspects of elastic and inelastic diffraction in Sec.\ref{ch:theory}. The results are presented and discussed in Sec.\ref{ch:result}.

\section{Grazing incidence fast atom diffraction (GIFAD)\label{ch:setup}}

The grazing incidence fast atom diffraction at crystal surfaces uses atoms in the keV energy range as probed with incidence angles  $\theta_i$ around 1 deg.
so that the full diffraction pattern can be recorded in one take on a position-sensitive detector \cite{Morosov_1996,Lapington_2004,Lupone_2015,Lupone_2018} as sketched in Fig.\ref{fgr:fig1}.
A commercial ion source delivers ions at the desired energy, they pass inside a charge exchange cell filled with the same gas, where a significant fraction is neutralized by resonant electron capture, see \textit{e.g.} Ref.\cite{Debiossac_PCCP_2021}.
After this cell, the ion fraction is deflected away and the spatial extent and angular divergence of the neutral beam is controlled by two co-linear diaphragms adjustable between 20 and 100 $\mu$m, separated by a distance close to half a meter before entering into the UHV chamber with the target.
If the projectile encounters a large enough terrace, it undergoes quasi-specular reflection and the projectiles are scattered within a cone with an opening angle of $\theta_i$.
Since keV atoms are easily detected and imaged by micro-channel plates, GIFAD was able to record a few images per second \cite{Atkinson_2014,Debiossac_2017} with an old ion source.

GIFAD offers a high topological resolution of a few $pm$ on atomic structure, \textit{e.g.} surface rumpling \cite{Schuller_rumpling} or charge transfer \cite{Khemliche_2009}, simple semi-quantitative interpretation \cite{Debiossac_PCCP_2021} and, when compared with exact quantum scattering code \cite{Rousseau_2007,Aigner_2008,Zugarramurdi_2012,Diaz_2016b}, a parameter free accuracy \cite{Debiossac_PRB_2014, Debiossac_JPCL_2020}.
The temperature of the surface affects both its nuclear and electronic systems.
We use the well-documented system of helium on LiF(001), where the large band-gap prevents electronic contributions allowing interpretations of inelastic effects only in terms of thermal motion of the surface atoms.
%
%we investigate the effect of temperature with a test crystal, LiF(001), a large band-gap insulator where electronic excitation's can be kept at a negligible level \cite{Auth_1997,Roncin_1999} so that all observed modifications can be traced to thermal displacement of surface atoms.  
%
\begin{figure}\includegraphics[width=0.95\linewidth]{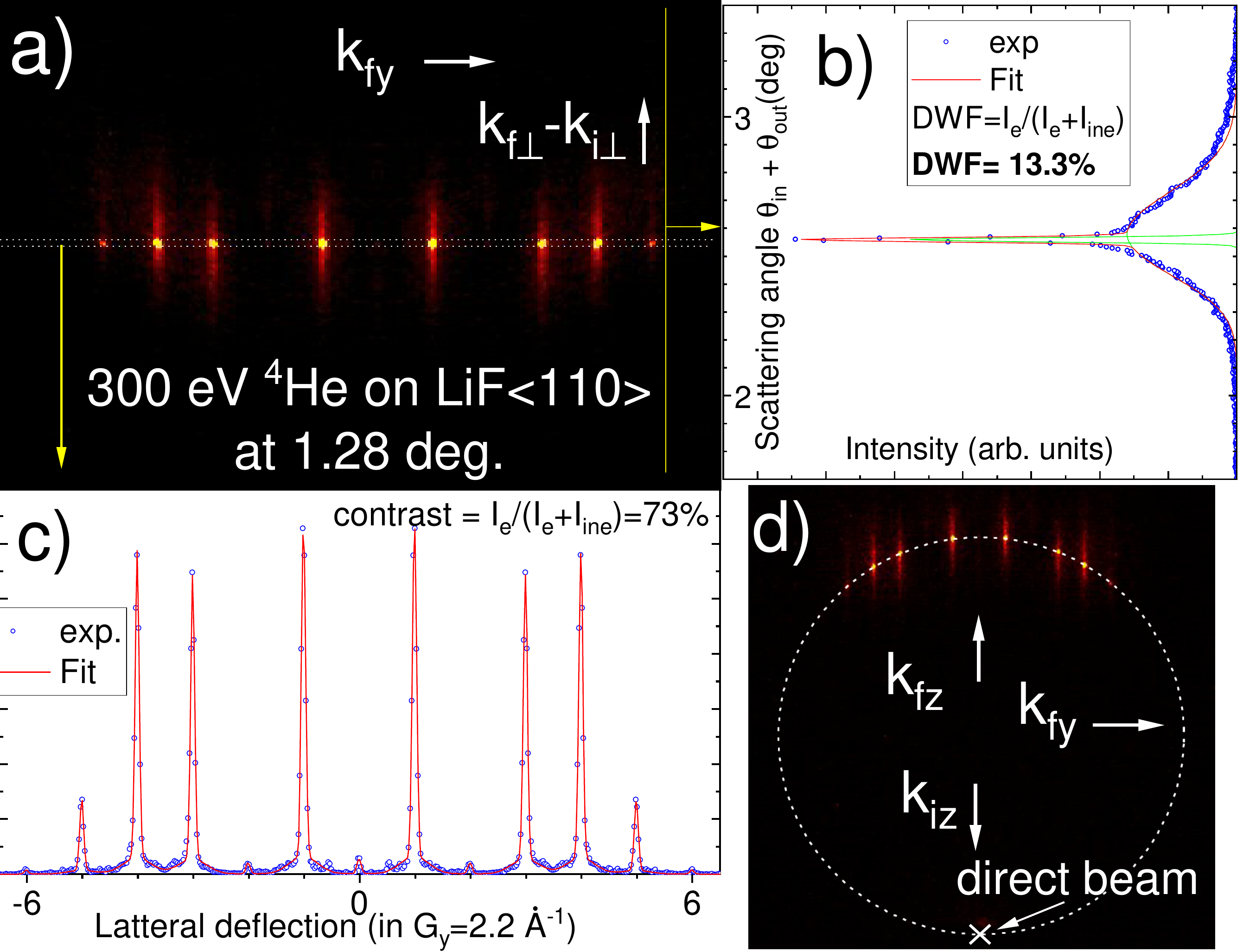}
	\caption{a) quasi polar transform of the raw diffraction image in panel d). The polar scattering profile b) corresponds to a full projection of a) onto the vertical axis. It is fitted by the sum of a narrow Gaussian and a broad log-normal profile used to evaluate the DWF=$I_e/I_{tot}$ with $I_{tot}=I_e+I_{ine}$. Panel c) corresponds to the intensity in a narrow horizontal band centered at the specular angle. The contrast measured on the Laue circle (c) is 73\% much larger than the DWF=13\%.
	\label{fgr:He_transform}}
\end{figure}

\begin{figure}
	\includegraphics[width=0.95\linewidth,draft = false]{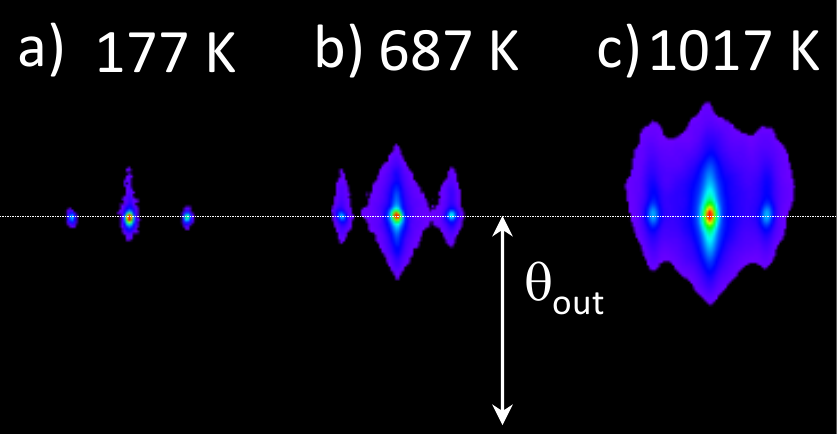}
	\caption{Three diffraction images recorded with 500 eV helium impinging with $\theta_{in}$=0.75 deg. on LiF [100] at temperatures of 177 K a), 687 K b) and 1017 K c). The images are normalized to the maximum intensity corresponding here to the elastic specular spot. The rainbow color palettes are identical with a threshold at 3\% of the maximum intensity. \label{fgr:three_temp}}
\end{figure}
A definition of the $(x,y,z)$ axis is displayed in Fig.\ref{fgr:fig1} together with a typical raw diffraction image on GaAs at elevated temperatures. 
Another raw diffraction image is  plotted in Fig.\ref{fgr:He_transform}d) for a LiF crystal and a helium beam oriented along the [110] direction.
These images correspond to a direct mapping of the final velocity or wave vector ($k_{fy},k_{fz}$) of the scattered projectile perpendicular to the crystal axis. 
Bright elastic diffraction spots are clearly visible and located on a single circle corresponding to energy conservation of the motion in the ($y,z$) plane perpendicular to the crystal axis probed : $k_{fy}^2 + k_{fz}^2 = k_\perp^2 = cst$.
A polar-like transformation \cite{Debiossac_Nim_2016} brings this Laue circle into a straight line displayed in  (Fig.\ref{fgr:He_transform}a)).
%The Laue circle where the bright spots associated with elastic diffraction are located is transformed in a straight line (Fig.\ref{fgr:He_transform}a)) using a polar-like transform \cite{Debiossac_Nim_2016}.
The intensity in this narrow stripe of maximum intensity is reported on Fig.\ref{fgr:He_transform}c) and shows well-resolved diffraction peaks equally separated by multiples of the Bragg angle $\theta_B=\arctan(G_y/\hslash k_{//})$, where $\hslash k_{//}= \hslash k\cos{\theta_{in}}$ is the projectile momentum parallel to the crystal axis.
$G_y=2\pi/a_y$ is the reciprocal lattice vector associated with the distance $a_y$ between atomic rows perpendicular to the probed crystal axis, taken here as the $x$ direction.
To derive the structure factor, only the elastic intensities should be considered, however, when elastic intensity is significant, the elastic and inelastic relative intensities on the Laue circle $I_m$ were found identical \cite{Roncin_PRB_2017} so that Fig.\ref{fgr:He_transform}c) can be directly exploited.

The projection of Fig.\ref{fgr:He_transform}a) on the vertical axis produces the polar scattering profile in Fig.\ref{fgr:He_transform}b) showing a narrow elastic peak on top of a broader inelastic scattering profile.
The relative weight of the elastic peak can be estimated using a simple fit where the elastic component is represented by a narrow Gaussian peak and the inelastic one by a broader, slightly asymmetric log-normal profile $f(\theta)=\frac{1}{\text{w} \theta\sqrt{2\pi}} \exp(-\frac{(\ln{\theta}-\ln{\theta_s})^2}{2\text{w}^2})$ \cite{lognormal,Manson_PRB_2008,Pan_Polar_2021}.
Assuming that the image contains only the gently scattered projectile that did not encounter major surface defects, this ratio DWF=$I_e/I_{tot}$ is believed to be a direct measurement of the Debye-Waller factor. 
It is different from the standard definition used in TEAS DWF=$I_e/I_0$ where $I_0$ would be the intensity of the direct beam which, in practice, is never measured directly with the same detector.
In GIFAD, the direct beam is always measured, either directly or through a calibrated attenuation grid, because both the exact beam position and line-profile are mandatory for accurate analysis.
It reveals that for the present cleaved LiF sample, the intensity scattered by the surface drops rapidly from close to unity an incidence angle above to 2-3 deg down-to only a few percent below 0.5 deg.
This reduced reflectively is in part due to the overlap with the sample surface but is probably mainly due to the distribution of terrace height resulting from the cleavage.

\begin{figure}
	\includegraphics[width=1\linewidth,draft = false]{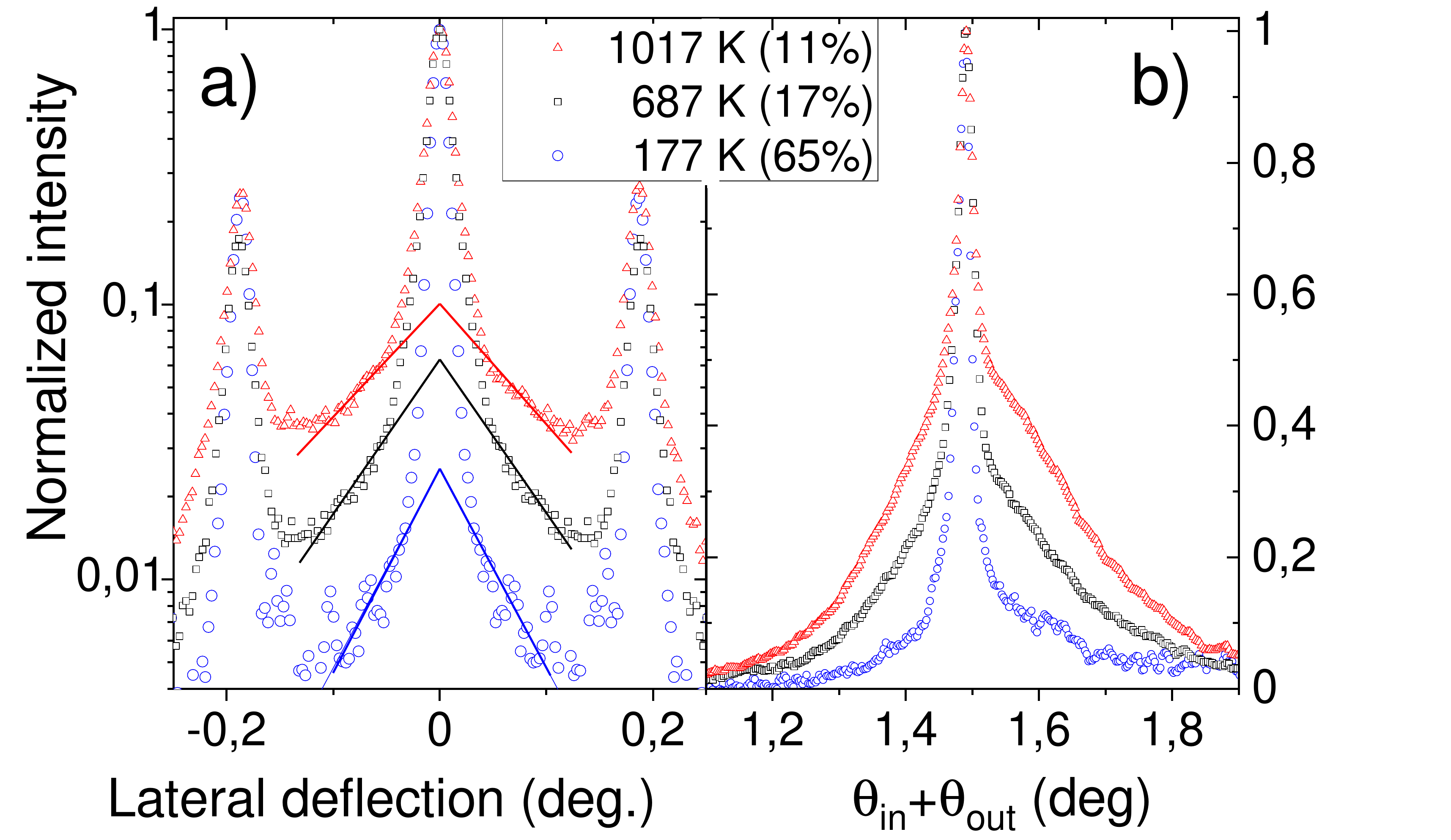}
	\caption{For the three images in Fig\ref{fgr:three_temp}, panel a): lateral profiles on the Laue circle, the lines are here to outline the exponential decay of the peak tail. Panel b): polar scattering profiles.\label{fgr:Profiles}}
\end{figure}

Fig.\ref{fgr:three_temp} shows three diffraction images recorded at temperatures of 177K, 687K, and 1017K measured with a type-N thermocouple mechanically pressed on the backside of the Omicron sample plate.
The transition from a spotty pattern to a much more diffuse one is clearly visible and is discussed in detail along the vertical and horizontal axis corresponding respectively to the polar profile and lateral deflection.
%Before considering quantitatively,  Fig.\ref{fgr:Profiles} reports the associated horizontal intensity at the specular angle and the projected vertical intensity corresponding to the polar scattering profiles.
On Fig.\ref{fgr:Profiles} a) the intensity on the Laue line is displayed in the log scale.
Within experimental uncertainty due to slightly different beam conditions, the narrow elastic peaks do not change shape but the inelastic contribution increases both in intensity and in width as outlined by the full lines.% suggesting an exponential decay of the inelastic component with the lateral deflection from the elastic position.
More precisely, in this example, both the extrapolated inelastic peak intensity and the exponential decay range increase quasi-linearly with the temperature.

Fig.\ref{fgr:Profiles}b) reports the polar scattering profiles of Fig.\ref{fgr:three_temp}. %a), b) and c)
The broad inelastic profile is also clearly identified with a relative height and a width growing with temperature so that the relative elastic intensity decay rapidly.% as reported in Fig.\ref{fgr:DWF&Vis}.

At this stage, it is useful to compare with the first investigation of the temperature dependence performed when elastic diffraction was not yet demonstrated and where all peak shapes seemed to depend on temperature \cite{Khemliche_2009,Busch_2012,Debiossac_PRB_2014}, probably because of a limited surface coherence.
In this context, the Debye-Waller factor was tentatively attributed to the ratio of the narrow peak relative to the total intensity observed on the Laue circle\cite{Rousseau_2008,Bundaleski_2011,erratum}, \textit{e.g.} in Fig.\ref{fgr:He_transform}b) or  Fig.\ref{fgr:Profiles}a), rather than from the projected polar profile in Fig.\ref{fgr:He_transform}c) or Fig.\ref{fgr:Profiles}b).
This point of view, obviously underestimates the overall inelastic intensity and strongly depends on the primary beam profile, and is not used anymore.

%The figure \ref{fgr:DWF&Vis} displays both approaches outlining the optimistic point of view of Fig.\ref{fgr:Profiles} a), however both indicate a exponential decay with the cube of the angle of incidence.
%\begin{figure}
%	\includegraphics[width=0.9\linewidth,draft = false]{DWF&Vis.pdf}
%	\caption{In ref\cite{Rousseau_2008}, the DWF was erroneously attributed to the intensity scattered on the Laue circle where %the elastic intensity is located. The , corresponding to\label{fgr:DWF&Vis}}
%\end{figure}

\section{$\phi$-scan, $\theta$-scan, E-scan and T-scan \label{ch:phi_scan}}
A typical GIFAD experiment begins with the introduction or preparation of a new sample surface and subsequent annealing.
The first measurements consist of a rapid azimuthal scan, $\phi$-scan, where the target surface is rotated in-plane (around the $z$ axis in Fig.\ref{fgr:fig1}) in order to identify its crystal axis. 
It does not require that diffraction is observed because a simple analysis of the width of scattered lateral profiles; $\langle k_{fy}^2 \rangle^{1/2}$ in Fig.\ref{fgr:He_transform}a, \ref{fgr:He_transform}c, \ref{fgr:He_transform}d, \ref{fgr:Profiles}a), as a function of the target azimuthal angle $\phi$ is enough to identify the principal crystal axis.
The procedure called triangulation \cite{Feiten_2015,Kalashnyk_2016,Debiossac_PCCP_2021} is used where the angle of incidence $\theta$ ($ \equiv k_{iz}$) is kept constant.
In practice, measuring the relative width $\langle k_{fy}^2 \rangle^{1/2} /k_{iz}$ compensates for a minor tilt between the surface normal and the rotation axis \cite{Kalashnyk_2016,Sereno_2016}. 
After proper alignment on the desired axis, the surface is prepared by various methods until a good diffraction pattern is observed.
Here, we used LiF single crystal previously irradiated by $\gamma$-rays \cite{Gilman_1958} giving a pronounced yellow to orange color, and cleaved in the air just before introduction into the vacuum. Subsequent heating at 400 $^\circ$C during a few hours is usually enough to recombine the color centers and to record clean diffraction images with well-resolved elastic spots.
The size of the diffraction spots provides indications of the surface coherence length.
The lateral angular resolution corresponding to $~$1/10 of the Bragg angle indicating a transverse coherence length $\delta_y=(k\delta\theta)^{-1} \sim10~a$ with $a$ the lattice unit. 
The equivalent vertical width suggests a longitudinal coherence length $\delta_z/\theta_{in}$ almost hundred times larger for $\theta=0.57$ deg so that $1/\theta=100$ (for circular diaphragms  $\delta k_z=\delta k_y$ so that $\delta_z=\delta_y$).
Thus coherent specular reflection requires a crystal without defect on a surface $\delta_S$ given by $\delta_S = \delta_y^2/\theta_{in}~$ $10^4$ $a^2$ equivalent to a tiny square hundred by hundred lattice units or $~10^5$ \AA$^2$.

In principle, a $T$-scan would consist of a simple variation of the target temperature leaving all other parameters unchanged.
Unfortunately, this is not compatible with the extreme sensitivity of grazing incidence.
In GIFAD, the target surface is easily positioned within 10 to 20 $\mu$m from the beam: when it is not  inserted enough, the primary beam is still present on the image, whereas when it is inserted too much, the direct beam impacts the edge of the $\approx 1$ mm thick crystal and even the scattered beam disappears from the detector.
However, thermal expansion of the target crystal and of the manipulator induce much larger displacements as well as minor angular tilts producing major effects in GIFAD (see \textit{e.g.} the $\theta^3$ dependence in Fig.\ref{fgr:DWF_Et3}).
For this reason, we were not able to record the three images displayed in Fig.\ref{fgr:three_temp} one after the other.
Instead, we had to wait for a stable temperature before realigning the target and restoring a comparable incidence angle. In the following, we rationalize this approach by performing several $\theta$-scan or $E$-scan at different temperatures.
From these variations, we interpolate between measured angular values, for instance with a B-spline, to restore a temperature variation.
It should be noted that at elevated temperatures, typically above 800 K, particles emitted from the heating filament produce an additional noise appears on our detector that may ruin a temperature variation whereas, taking time to find a stable detector bias and data-acquisition-mode can reduce the noise level allowing a more stable $\theta$-scan or $E$-scan.

\section{elastic and inelastic diffraction \label{ch:theory}}

For ideal crystalline surfaces with atoms at the equilibrium position, it has been shown that the rapid movement parallel to the crystal axis ($x$) is decoupled from the slow one in the ($y,z$) perpendicular plane \cite{Zugarramurdi_2012,Zugarramurdi_NIM2013,Farias_2004,Debiossac_PRA_2014,Diaz_2016b,Danailov_1997}. 
Therefore, elastic diffraction of fast atoms along a crystal axis is equivalent to that of a much slower particle with an energy $E_\perp = E \sin^2 \theta$ evolving in the averaged potential $V_{2D}(y,z)=\int_x V_{3D}(x,y,z)$. 
Experimentally, this axial channeling approximation (ASCA) results in diffraction taking place only in the ($y,z$) perpendicular plane \cite{Farias_2004,Winter_PSS_2011,Debiossac_PCCP_2021}  (see also \cite{Busch_2012, Zugarramurdi_NIM2013} for the limitations).
From a spatial point of view, the elastic diffraction can be seen as the coherent part of the scattered waves and this later can only build up at a location corresponding to the equilibrium position, the one of the center of the vibrational wavefunction.
In contrast, inelastic diffraction corresponds to a situation where momentum and/or energy has been exchanged with the surface \textit{i.e.} with the vibration of the surface atoms or phonons, breaking the exact translation symmetry of the ideal surface and requiring \textit{a priori} to abandon the ASCA \cite{Frisco_2019,Frisco_2020}.

In TEAS, x-ray or neutron diffraction, this situation is described by the Debye-Waller factor DWF=$I_e/I_0 = e^{-\sigma_\phi^2}$ where $\sigma_\phi^2$ is the variance of the phase distribution induced the thermal displacement $\sigma_z$ of surface atoms. 
In a Debye model of solids, each atom is described by a local harmonic oscillator with frequency $\omega_D$ and the thermal amplitude is Gaussian $\sigma_z(T)$ resulting in a phase coherence $e^{-\sigma_\phi^2}=e^{-q^2\sigma_z^2}$ where $q$ is the projectile momentum.
It describes how the elastic diffraction intensity progressively vanishes when the amplitude $\sigma_z$ of the thermal disorder becomes comparable to the projectile wavelength $\lambda=h/q$.
From a momentum point of view, the DWF can be written as $e^{-E_r/\hslash\omega}$ where $E_r = q^2/2m$ is the binary recoil energy, which can be interpreted as the Lamb-Dicke probability of recoil-less emission that the harmonic oscillator exchanges the momentum $2q$ without changing its energy level.
This DWF does not apply directly to GIFAD, even when considering $\lambda_\perp=\lambda/ \sin\theta_{in} \gg \lambda$, because the momentum exchange of $2k\sin\theta_{in}$ needed for specular reflection of the primary beam is acquired via several successive fast gentle collisions with surface atoms.
With a rigid lattice model and an exponential repulsive mean planar potential of stiffness $\Gamma$, $V(z)=V_0 e^{-\Gamma z}$, the trajectory is analytic and the binary momentum exchanges can be calculated to obtain the classical energy loss $\Delta E_{Cl}$ in eq.\ref{eq:DWF_Eloss} with $m_p$ the projectile mass and $a$ the lattice unit\cite{Manson_PRB_2008,Roncin_PRB_2017}.
The effective number $N=\frac{6}{\Gamma a \theta}$ of binary collision is defined such that $N$ times the individual deflection $d\theta$ is the specular deflection $2\theta_{in}$ and that $N$ times the individual binary recoil energy $E_r$ matches the analytic total energy loss $\Delta E_{Cl}=N~E_r$. The DWF for GIFAD writes:
\begin{equation}
DWF=\exp\left({-\frac{3\Delta E_{Cl}}{\hslash\omega_D} \coth{\frac{T_D}{2T}} }\right), ~ \Delta E_{Cl} = \frac{2~m_p}{3~m} E \Gamma a \theta_{i}^3~\equiv N E_r
\label{eq:DWF_Eloss}        
\end{equation}
For the He-LiF system, the product $\Gamma a$ is close to 14 \cite{Debiossac_JPCL_2020,Pan_Polar_2021} so that $N\approx 2/\theta$ can be large explaining why elastic diffraction could be observed with $E_\perp$ close to one eV \cite{Winter_PSS_2011, Debiossac_PCCP_2021} whereas TEAS is usually limited below 100 meV. 
Alternately, with this reduced decoherence, GIFAD can explore the higher temperatures that are needed for MBE.

This naive, purely repulsive description was improved by taking into account an attractive part of the potential, for instance, van der Waals contributions \cite{Beeby_1971,Bocan_PRA_2016,Cueto_2016,Gravielle_2019}, responsible for the physisorption well of depth $D$. In elastic diffraction, the effect of such attraction is the presence of bound-state resonances \cite{Jardine_2004,Debiossac_PRL_2014} and the increase of the rainbow angle at low energy \cite{Debiossac_JPCL_2020}. These are naturally accounted for using a quantum approach \cite{Debiossac_PRL_2014,Debiossac_JPCL_2020} or modeled in semi-classical \cite{Bocan_2021} or optical method such as the hard corrugated wall by the Beeby correction indicating that the effective impact energy $E_\perp$ increases to $E_\perp +D$ \cite{Beeby_1971,Taleb_2019,Debiossac_PRB_2016} or considering a modified angle of incidence $\theta_\text{eff}=\sqrt{\theta_{in}^2 + D/E}$.
The Beeby correction also decreases significantly the DWF in TEAS \cite{Vidali_1988}.
In GIFAD, we also found that the mere presence of a tiny well significantly modifies the stiffness of the potential by bringing the turning point much closer to the surface plane \cite{Pan_Polar_2021}. 
This can be expressed quantitatively using a Morse potential $V_M(z)=D(e^{-\Gamma(z-z_0)}-2e^{-\Gamma/2(z-z_0)})$ and looking for the turning point $z_t$ where $V_M(z_t)=E_{\perp}$ the effective stiffness is :
\begin{equation}
\Gamma_\text{eff}(E_\perp)= \Gamma\left[1+\left(1+\frac{E_\perp}{D}\right)^{-1/2}\right]
\label{eq:GammaEff}\end{equation}
\begin{figure}\includegraphics[width=0.9\linewidth]{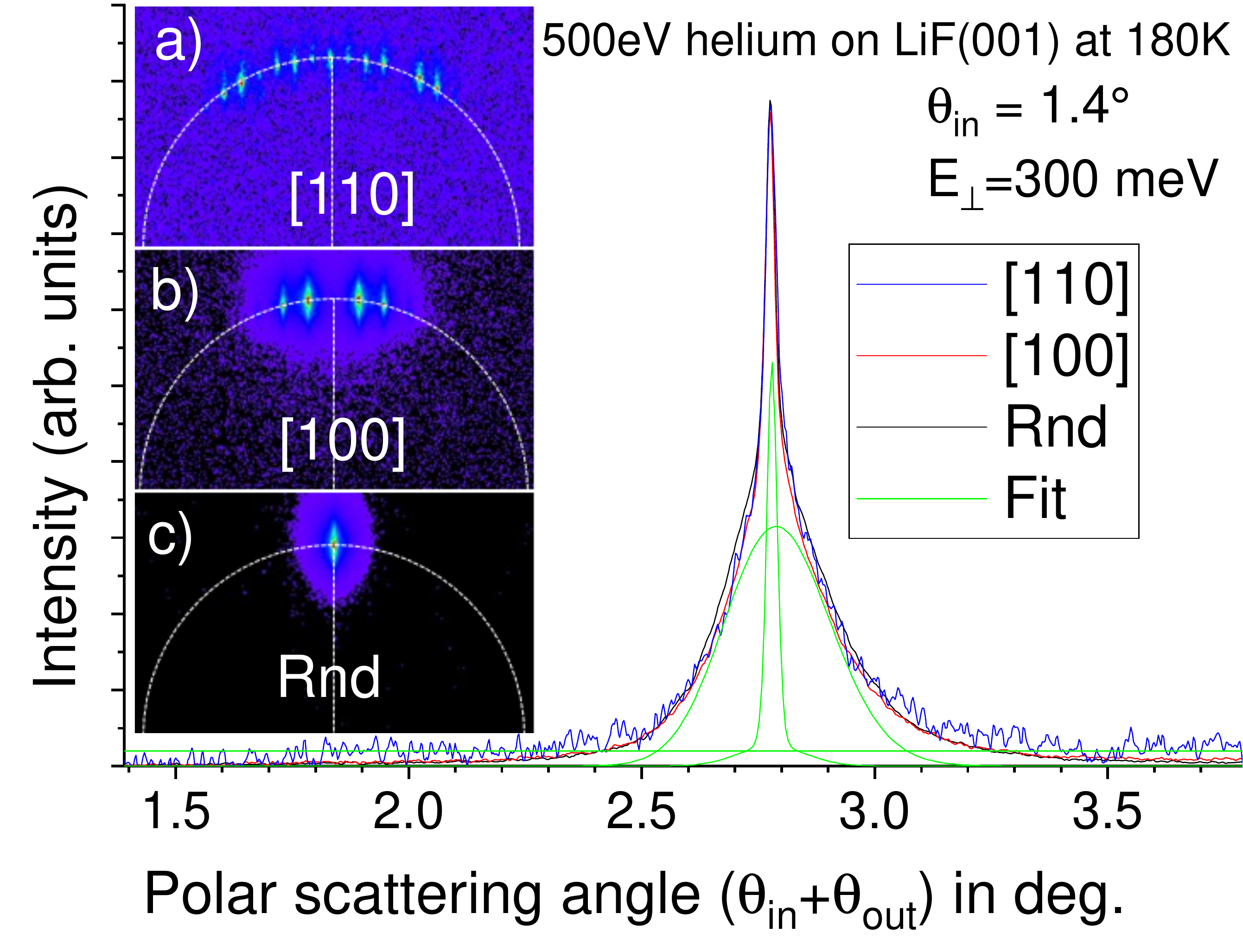}
	\caption{a), b) and c) are raw diffraction images of 500 eV helium incident at 1.4$^\circ$ on LiF at 180 K oriented along the [110], [100], and
random direction, respectively (in part, taken from \cite{Pan_Polar_2021}). 
The resulting polar scattering profiles are almost identical, with a narrow elastic peak at $\theta_{out}=\theta_{in}$ on top of a broader peak fitted by a log-normal profile \cite{lognormal} (green curves).\label{fgr:Scaled_polar}}
\end{figure}

This increased stiffness was already identified in TEAS \cite{Rieder_1982} but it has much less consequence since at normal incidence the projectile hits a single surface atom, only the time scale $\tau \approx \Gamma/v_\perp$ depends on $\Gamma$ not the magnitude of the exchanged momentum $2q$ and therefore not the coherence ratio.
It is the reverse in grazing scattering, for identical values of $E_\perp$ the time scale $\tau$ for bouncing from the surface are identical in TEAS or GIFAD, but the time needed for a single quasi binary collision is now $\tau'\approx v_{//}/a =\tau/N$ independent on $\Gamma_\text{eff}$, while its magnitude $\approx 2q/N$ depends directly on the effective stiffness $\Gamma_\text{eff}$. 
In summary, for GIFAD, the stiffness $\Gamma_\text{eff}$ governs the momentum transferred in each collision, a stiffer interaction potential needs fewer collisions for specular reflection and each of them becomes more violent leading to an overall reduction of the DWF.

At the atomic level, the temperature is modeled by the spatial extend of a surface atom of mass $m$, which, in a Debye harmonic model is Gaussian profile with $\sigma_z^2 (T)=\frac{3\hslash}{2m\omega}\coth{\frac{T_D}{2T}}\sim\frac{3\hslash}{2m\omega}\frac{2T}{T_D}$, where $T_D$ is the Debye temperature such that $\hslash\omega =k_BT_D$ with $k_B$ the Boltzmann constant.

\section{results \label{ch:result}}
%We first report the experimental results where the measured DWF \textit{i.e.}  the coherence ratio or the elastic scattering ratio appears stable.
%Then the results are presented for several $\theta_{scan}$ at different temperatures from which we try to build an interpolated $T_{scan}$ or $T$ dependence's in sub-section \ref{ch:T_scan}. The width of the scattering profiles and their evolution with the surface temperature is examined immediately after.

\subsection{Orientation of the surface}
When investigating the polar scattering profile \cite{Pan_Polar_2021}, it was shown that the shape of the polar scattering profile does not depend significantly on the orientation of the surface. This is illustrated in Fig.\ref{fgr:Scaled_polar} and we have checked that this similarity remains valid at different temperatures.
Since both the data acquisition and the data analysis are simpler for random orientation of the surface where only one specular order ($m=0$) is present, most of the temperature variations presented here correspond to this crystal orientation. The other data were recorded along the [100] direction where the corrugation amplitude is reduced generating fewer diffraction orders as visible when comparing Fig. \ref{fgr:three_temp} with Fig.\ref{fgr:He_transform}.

\begin{figure}
	\includegraphics[width=0.80\linewidth,draft = false]{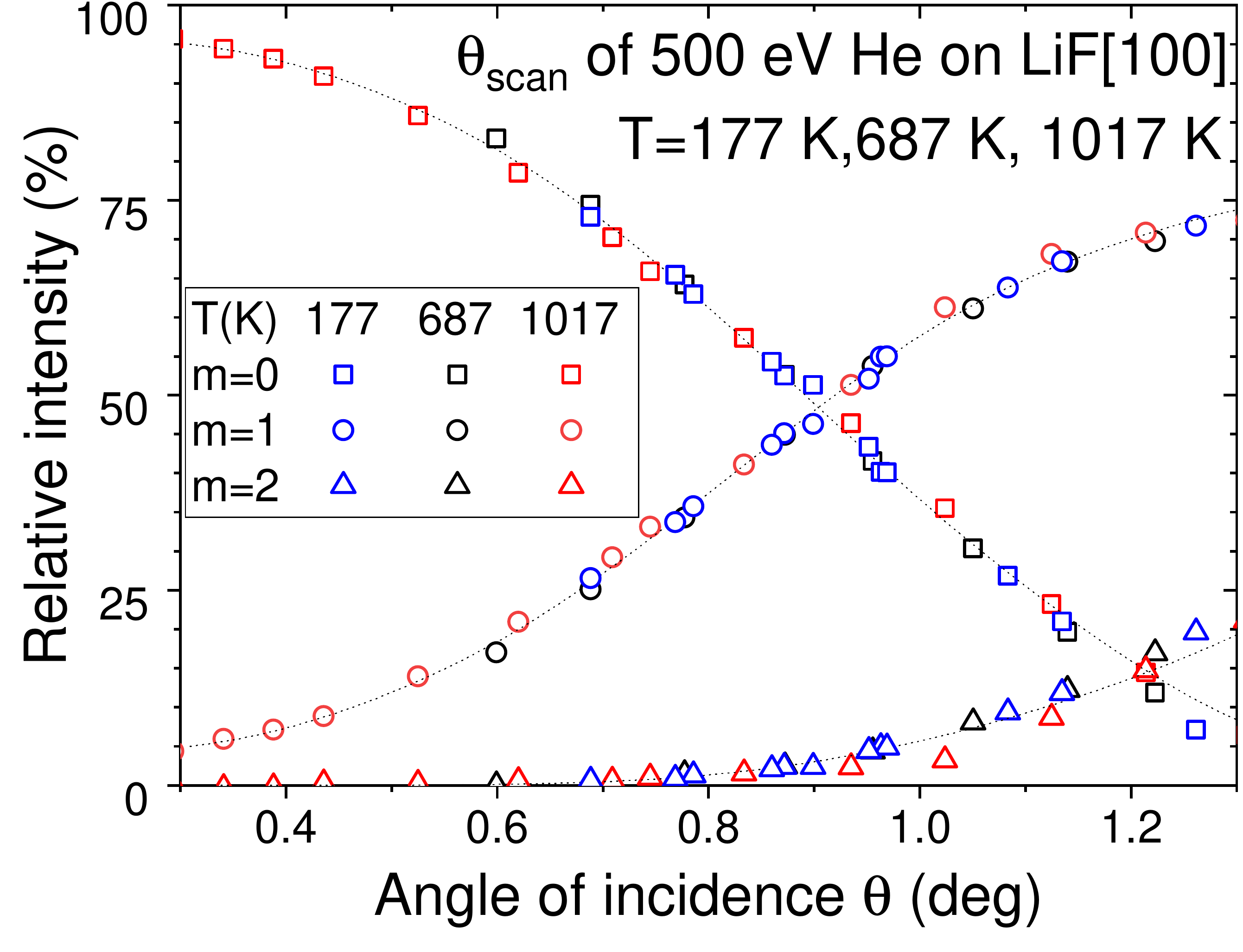}
	\caption{The relative intensities $I_m$ of (\myholowcircle{black})$m=0$, (\myholowsquare{black})$m=\pm1$ and ($\triangle$)$m=\pm2$ extracted from the elastic intensities recorded in three $\theta$-scan performed at temperatures of \myholowcircle{blue} 177 K, \myholowcircle{black} 687 K, \myholowcircle{red} 1017 K, fall on top of each other. \label{fgr:Im_de_T}}
\end{figure}

\begin{figure}\includegraphics[width=0.80\linewidth]{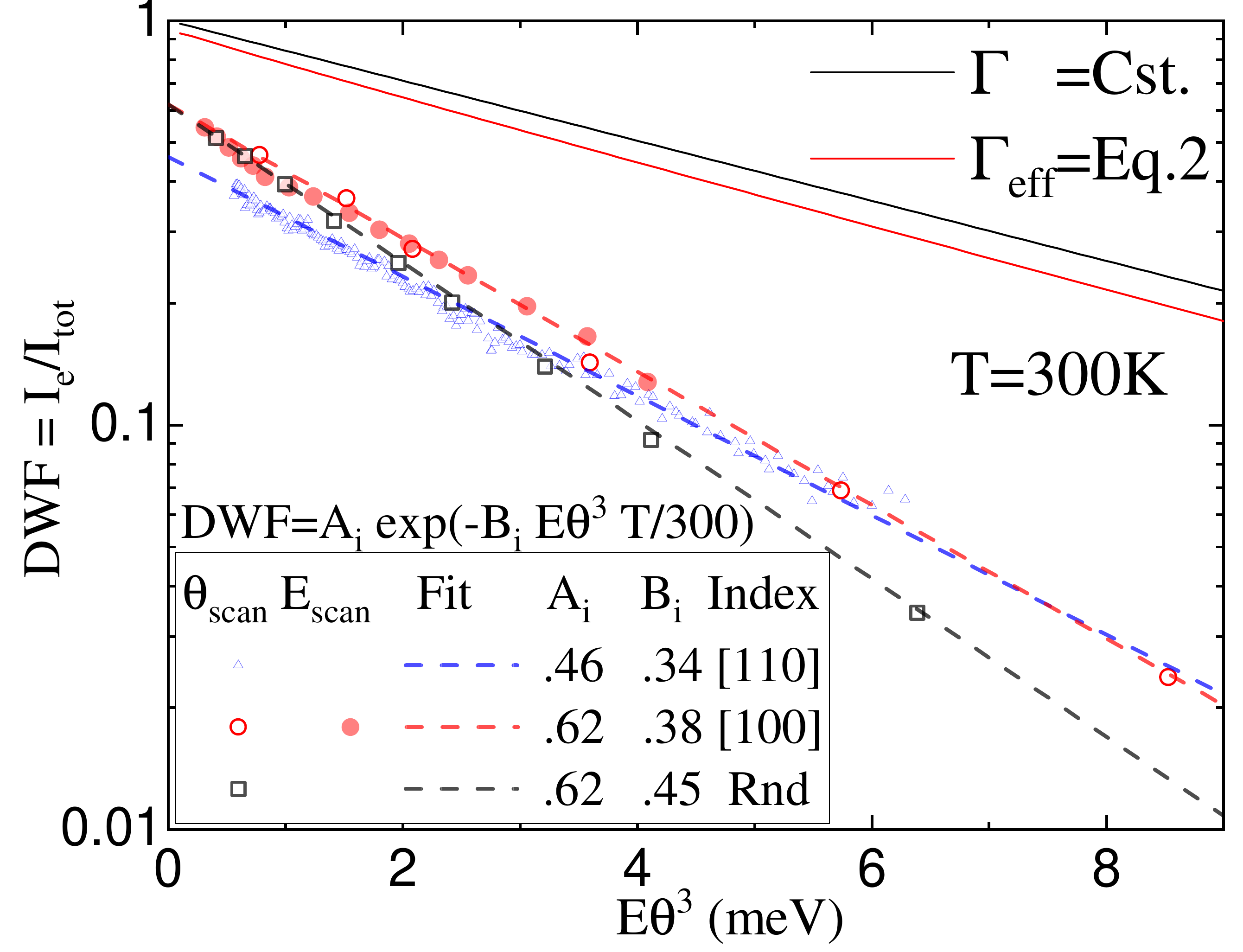}
\caption{The DWF=$\frac{I_e}{I_t}$ from 3 LiF samples are reported as a function of $E \theta^3$. The dotted lines are fits to the data while the solid lines are numerical prediction of eq.\ref{eq:DWF_Eloss} with (red A=.94,B=.19) and without corrections (black A=1,B=.17) due to the attraction towards the surface \label{fgr:DWF_Et3}}
\end{figure}

\subsection{Bragg angle and form factor}
The main focus of this paper is the temperature dependence of the elastic ratio which is presumed to depend mainly on the thermal movement of the surface atoms.
%In addition the surface undergoes thermal expansion and the shape of its electron density, as probed by GIFAD as a form factor could also change. 
When analyzing the intensity on the Laue circle as displayed in Fig.\ref{fgr:He_transform}c), we can extract the value of the Bragg angle as well as the relative intensities $I_m$ associated with each diffraction order $m$. 
Both could change with the temperature due to thermal expansion and possible reconstruction of the surface.
This could be investigated in detail but we have only checked that the surface equilibrium positions and GIFAD form factor do not change significantly. 
The evolution of the Bragg angle is compatible with the thermal expansion coefficient measured by TEAS \cite{Ekinci_2004} and the Fig.\ref{fgr:Im_de_T} shows that in spite of important variation of the scattering profiles visible in Fig. \ref{fgr:three_temp} and \ref{fgr:Profiles}, the relative intensities $I_m$ measured along the [100] direction do not change significantly with temperature.
This aspect is not discussed further because, in practice, it is by-passed when analyzing the random direction where the reciprocal lattice vector is absent.

%\begin{figure}
%	\includegraphics[width=0.95\linewidth,draft = false]{Bragg_angle.pdf}
%	\caption{A diffraction pattern is fitted by seven equally spaced peaks with identical shapes. The inset shows the evolution of the $\chi_i^2$ for the Bragg angle suggesting an ultimate accuracy of a few $10^{-4}$ deg, much below the peak fwhm of 0.015 deg. almost four time to the full size of the inset. \label{fgr:Bragg_angle}}
%\end{figure}
\begin{SCfigure*}[][!t] %SCfigure*
%\begin{minipage}[t][5cm][b]{1\textwidth}
 \centering
 \includegraphics[scale=0.49,draft = false]{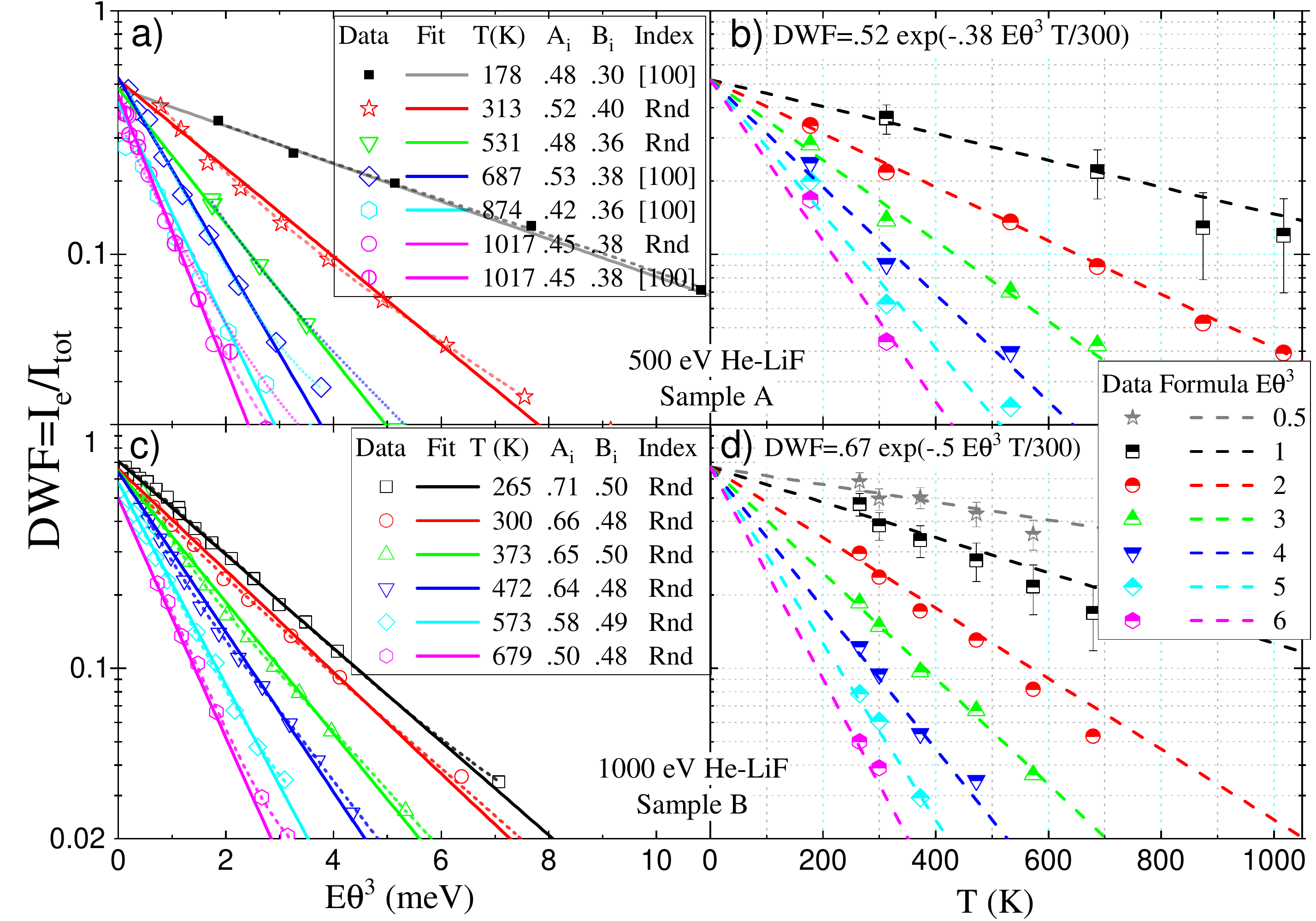}  
 \caption{Panels a) c), report the DWF as a function of the reduced parameter $E\theta^3$ for two different LiF samples. The dotted lines are B-spline interpolation used to derive the temperature dependence at fixed values of $E\theta^3$ in b) d) and reproduced by a global formula. At very low angles of incidence, corresponding to $E \theta^3 \leqslant$ 1 meV, where a diffuse scattering background is present an error bar estimated to 5\% is plotted. 
 \newline \newline \newline \newline \newline
 }
  \label{fgr:DWF_AB}
\end{SCfigure*}

\subsection{Sample quality}
All our samples have been prepared by cleaving at air LiF crystals previously irradiated by $\gamma$ rays \cite{Gilman_1958}, however, the Debye-Waller factor varies from sample to sample and also depend on the actual part of the surface illuminated by the atomic beam. 
As a worst case, we experienced a variation up to almost a factor two, after changing the target orientation and position when switching from Random to [100] direction. 
Exploring a few target positions is usually enough to optimize the DWF. 
We also observed a slow degradation of the diffraction images with time even at a few $10^{-10}$ mb pressure.
When taking a sample that was left in a vacuum for a few weeks, the measured DWF is systematically lower, even after a short thermal treatment.
This is illustrated in Fig.\ref{fgr:DWF_Et3} where three sets of data recorded on different samples are reported.
In spite of a scattering of data points in each set, the data follow different curves and therefore different DWF, but the trends are similar. In the final analysis, we have kept only two samples with large enough temperature variations and we provide the results as separate sets.

\subsection{Energy and angular variations}

As detailed in section \ref{ch:theory}, the effective DWF adapted to GIFAD is expected to scale with $\exp(-E\theta_{in}^3)$ where $E$ is the primary beam energy and $\theta_{in}$ the angle of incidence.
This was suggested from the spatial approach \cite{Rousseau_2008} considering a that the scattering by a row of $N$ atoms should have a mean thermal amplitude $\langle\sigma_z^2\rangle=\sigma_z^2 /N$ or from momentum transfer along the trajectory \cite{Manson_PRB_2008,Roncin_PRB_2017} because the classical recoil energy loss is also expected to scale with $E\theta_{in}^3$.
This dependence was first observed at room temperature by reporting DWF measured on a large set of energy and incidence angles \cite{Pan_Polar_2021}.
The Fig.\ref{fgr:DWF_AB}a) and \ref{fgr:DWF_AB}c) display similar dependence's recorded during E-scan and $\theta$-scan at different temperatures. 
All curves indicate a pronounced exponential decay illustrated by the straight lines resulting from independent fits $DWF=A_i\,\exp(-B_i\,E\theta^3\,T/300)$ where both $A_i$ and $B_i$ were left free and without any weighting of the data.

%\FloatBarrier
\subsection{Temperature dependence \label{ch:T_scan}}

According to the announced strategy, we do not use the fits of Fig.\ref{fgr:DWF_AB}a) to evaluate the temperature dependence but we use the B-spline interpolation of the data, plotted as dotted lines Fig.\ref{fgr:DWF_AB}a) to produce the temperature dependence displayed in Fig.\ref{fgr:DWF_AB}b) at fixed values of $E\theta^3$ between 0.5 meV and 6 meV.
Here again, the exponential character is readily visible. 
In contrast with Fig.\ref{fgr:DWF_AB}a), we now try to apply a unique formula to describe all the data. 
The adjustment was performed by changing step-wise the parameters $A$ and $B$ until a decent visual impression is obtained.
In practice, we could have used the data from the fits through the $E\theta^3$ variation Fig.\ref{fgr:DWF_AB}a,c) but this would induce a possible bias forcing the final dependence.

In TEAS, the Beeby correction to the DWF alone is significant and limits the maximum coherence, it is often used to estimate the $D$ value (see e.g. \cite{Beeby_1971,Farias_2002,Tamtogl_2021}).
This is not the case in GIFAD, a numerical evaluation, reported as full lines in Fig.\ref{fgr:DWF_Et3} indicates that these corrections have a weak influence both on the  magnitude and the exponential range.

\begin{SCfigure*}[]
  \centering
\includegraphics[width=1.65\linewidth,trim={6mm 4cm 4mm 0},clip]{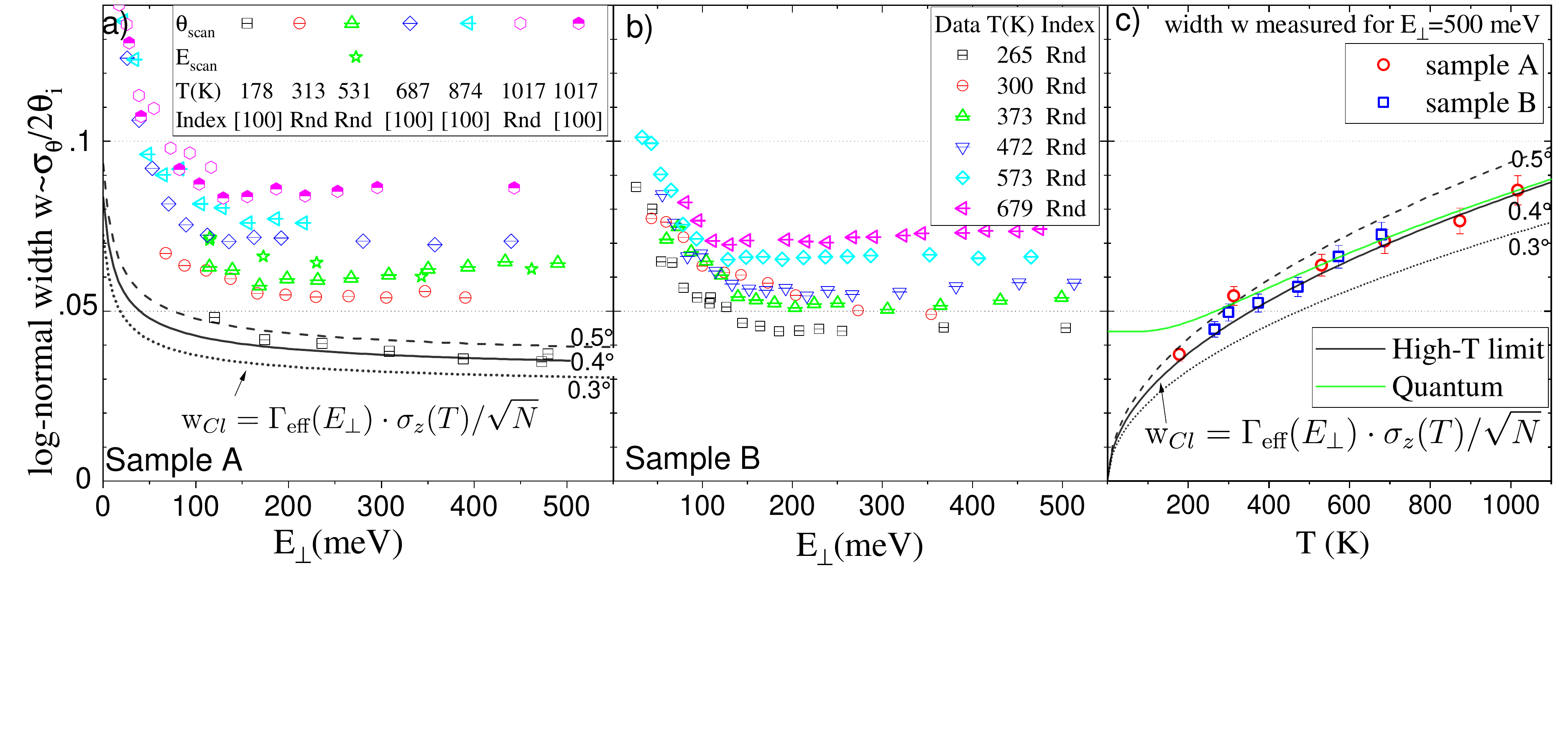}
  \caption{a) and b) inelastic scattering width $\text{w}$ as a function  of $E_\perp=E\theta^2$. The increase at low $E_\perp$ is mainly due to $\Gamma_\text{eff}(E_\perp)$. The $\text{w}$ measured at $E_\perp$=500 meV are reported in panel c) with the mean thermal amplitude $\sigma_z^2 (T)$. The lines in a) and c) correspond to Eq.\ref{eq:wCl} with different angles in the evaluation of $N$. (see text) The green curve marked as Quantum takes the zero point energy into account.}
  \label{fgr:wLN_AB_Ped}
 \end{SCfigure*}
 
\subsection{The inelastic scattering width}\label{ch:inelastic width}

The polar inelastic scattering profile is fitted log-normal form \cite{lognormal} and the relative width $\text{w}$ are reported in Fig.\ref{fgr:wLN_AB_Ped}a) and \ref{fgr:wLN_AB_Ped}b) as a function of the perpendicular energy $E_\perp$, a quantity that governs the distance of closest approach to the surface. 
%The widths are $<0.15$ so that $\text{w}$ is very close to the ratio  $\sigma_\theta/\theta_s$ where $\sigma_\theta$ is the standard deviation of the inelastic profile and $\theta_s$ the median value of the distribution, very close to the elastic scattering angle equal to $2\theta_{in}$.
The fact that this width was found \cite{Pan_Polar_2021} to depend mainly on $E_\perp=E\theta^2$ indicates that it is sensitive to the magnitude of the most violent inelastic collisions along the trajectory rather than to the integral effect of such collision which would be closer to $E\theta^3$.
Assuming that the inelastic collision is well-modeled by classical mechanics, the thermal motion $\sigma_z(T)$ of the surface atom induces, for each collision a log-normal scattering profile having a width $\text{dw}=\Gamma\sigma_z$ \cite{Manson_PRB_2008,Roncin_PRB_2017} or, equivalently a contribution to the angular straggling $d\sigma_\theta=\Gamma\sigma_z d\theta$ where $d\theta$ is the elastic deflection angle in this collision already estimated as $d\theta=(\theta_{in}+\theta_{out})/N$.
%because, for low value of $\text{w}$, $\text{w}\approx\sigma_\theta/\theta_s$ with $\theta_s=\theta_{in}+\theta_{out}$.
Adding the $N$ individual variances $d\sigma_\theta^2$ or using the mean log-normal width $\text{w}=\text{dw}/\sqrt{N}$ the classical scattering width is predicted: %$\text{w}_{Cl}=\Gamma\sigma_z/\sqrt{N}$.
\begin{equation}
\text{w}_{Cl}=\Gamma\sigma_z/\sqrt{N},\text{with } N=6/\Gamma a \theta_{in}
\label{eq:wCl}\end{equation}
%The comparison with experiment is tricky, the Eq.\ref{eq:wCl} reproduced the evolution during an $E$-scan where, neglecting for here the Beeby correction, $\theta_{in}$ is constant so that $N$ is also essentially constant but failed to fit the evolution during a $\theta$-scan while the experiment gives similar results during $E$-scan and $\theta$-scan \cite{Pan_Polar_2021}.
The comparison with experiment is tricky, the Eq.\ref{eq:wCl} reproduced the evolution during an $E$-scan but not during a $\theta$-scan while the experiment gives similar results during $E$-scan and $\theta$-scan \cite{Pan_Polar_2021}.
In this respect, the following discussion is only qualitative.
More precisely, during an $E$-scan where $\theta_i$ is fixed $N$ would stay constant so that, neglecting for here the Beeby correction, $\text{w}_{Cl}\propto \Gamma\sigma_z$ would remain constant.
However, as already observed at room temperature \cite{Pan_Polar_2021} and in Fig.\ref{fgr:wLN_AB_Ped}a) and \ref{fgr:wLN_AB_Ped}b) a sharp increase of $\text{w}$ is observed at low $E_\perp$. 
The agreement was established by taking into account the attractive forces \textit{i.e.} by replacing $\Gamma$ by $\Gamma_\text{eff}$ from Eq.\ref{eq:GammaEff}.
The increase at low energy could be then be attributed to the enhanced stiffness at low energy \cite{Pan_Polar_2021}.
The full, dashed and dotted lines indicate how the absolute values of $\text{w}_{Cl}$ depend on the angle of incidence of the hypothetical equivalent $E$-scan.
The \ref{fgr:wLN_AB_Ped}c) reports the value of $\text{w}$ measured at $E_\perp$ = 500 meV where is becomes stable. %The overall evolution with temperature seems to follow the expected dependence of $\sigma_z(T)$.
The lines using the same Eq.\ref{eq:wCl} now indicate that the evolution of the plateau values in Fig.\ref{fgr:wLN_AB_Ped}a) and \ref{fgr:wLN_AB_Ped}b) is compatible with the expected variation of $\sigma_z$ even if the low-temperature zero point motion (green curve) is not visible. 
The physical parameter $D$=8.5 meV \cite{Jardine_2004} and $T_D$=550 K \cite{Schuller_rumpling} correspond to well-accepted values in the literature and $\Gamma$=3.5 \AA$^{-1}$ was produced in a quantum calculation \cite{Debiossac_PRL_2014}.
It is close to the asymptotic value $\Gamma =2\sqrt{2W}=3.55 \pm 0.15$ \AA$^{-1}$ where $W = 12.2 \pm 0.5$ eV is the work-function of LiF.
Once again, the semi-quantitative agreement should be balanced by the fact that the model does not predict the observed similar behavior of $\text{w}(E_\perp)$ during an $E$-scan and a $\theta$-scan.

\section{Summary and Conclusion}
Using a definition based on the analysis of the polar scattering profile to isolate the elastic and inelastic components, the DWF can be evaluated for each diffraction image.
At comparable energy and incidence angle, the DWF is found independent of the crystal axis probed.
Due to the extreme sensitivity to mechanical deformations associated with temperature variations, the $T$-scans were performed indirectly via interpolation of $\theta$-scan and $E$-scan at different temperatures. 
At each temperature, the DWF specific to the multiple collision regime of GIFAD is shown to depend primarily on $E\theta^3$, differing from $E\theta^2$ in TEAS where a single collision regime prevails.
Within the present accuracy, a simple exponential decay is observed but, different LiF samples produce slightly different decay parameters and maximum coherence suggesting an important contribution of the defect density and terrace size distribution.
The effect of the attractive forces towards the surface have been investigated in TEAS.
It produces an increased impact energy, known as the Beeby correction, and an increased stiffness of the surface means planar potential energy surface due to a closer distance of approach towards the surface \cite{Rieder_1982}.
They also have the same consequences regarding the elastic diffracted intensities in GIFAD or TEAS but very different consequences in the inelastic behavior of GIFAD and TEAS.
%\cite{Taleb_2019,Debiossac_PRB_2016}
In TEAS, the Beeby correction is known to limit the maximum possible DWF \cite{Beeby_1971} while in GIFAD, the Eq.\ref{eq:DWF_Eloss} and Fig.\ref{fgr:DWF_Et3}c) indicate a limited influence on the maximum DWF.
As to the effective stiffness $\Gamma_\text{eff}$\cite{Rieder_1982}, it does not directly affect the DWF in TEAS, whereas, it enters the DWF factor in GIFAD because each binary collision becomes more violent as evidenced by the sharp increase of the inelastic scattering width $\text{w}$ at low values of $E\theta^2$ in Fig.\ref{fgr:wLN_AB_Ped}a),b).
The effect of $\Gamma_\text{eff}$ on the DWF, though larger than the Beeby correction is also reduced because it is, in part, balanced by the reduced number of collisions needed for specular reflection.  

With the LiF samples used here, the Fig.\ref{fgr:DWF_AB}b and d) indicate that GIFAD is not able to offer an internal temperature with an accuracy better than 20-50 $^\circ C$.
Combining consistently the width $\text{w}$ and the DWF could improve the accuracy but the sensitivity to sample quality appears as a severe limitation.
In contrast, if the main focus is to optimize the growth parameters to improve the surface quality in terms of coherence length, \textit{i.e.} mean distance $L_C$ between defects, GIFAD offers a unique handle with a very broad range of operation.
First, a simple $\phi$-scan \cite{Kalashnyk_2016} can identify crystallographic axis even without diffraction offering a first estimate of $L_C$ via the peak to background ratio of the $\phi$-scan \cite{Debiossac_PCCP_2021}.
When diffraction becomes visible, the presence of elastic diffraction, and its associated elastic peak width, readily gives insights on $L_C$.
Then, optimizing the DWF could give real time access to very large defect-free surfaces and the diagnostic is performed simultaneously on an illuminated surface $S\approx\oslash^2 /\theta$ on the order of 1 mm$^2$ for a diaphragm size of $\oslash$=100 $\mu$m.
This diagnostic is complementary to the elastic diffracted intensity which indicates in real time the detailed topology of the terminal layer.
We have shown here that the width $\text{w}$ of the scattering profile can be understood qualitatively in terms of a classical model using an effective stiffness $\Gamma_\text{eff}(E_\perp)$ and a thermal amplitude $\sigma_z(T)$.
This suggests that classical scattering simulation in grazing incidence, in general \cite{Danailov_2001} and in the context of GIFAD \cite{Frisco_2019,Frisco_2020} should produce a fair estimate of the inelastic profile.
However, \textit{In fine}, a quantum inelastic treatment as developed in TEAS \cite{Kraus_2015} and recent attempts to encompass both elastic and inelastic aspects under grazing incidence \cite{Schram_2018} should help connecting to the real world of surface phonons and their possible specific coupling to the multiple collision regime.

%Difficulty to produce very accurate inelastic data. This contrasts with the elastic diffraction which by nature have not been affected by defect otherwise they would appear in the inelastic regime.
%Either because the defect would modify the transferred momentum or simply change the phase 
%it seems that once a collision is inelastic, the width corresponds to that of a classical description where all individual binary collisions are inelastic.
%This naive conclusion is of course impaired by the simplified description of the surface atoms as trapped in a local Debye-oscillator. Some attempts have been 

%The inelastic width appears well-described by the classical scattering width indicating that, once inelastic all binary collisions should be treated inelastic.
%This suggests that classical scattering simulation as performed by Pfandzelter and, in the context of GIFAD by Gravielle should produce a fair estimate of the inelastic profile.

\section*{Conflicts of interest}
There are no conflicts to declare.

\section*{Acknowledgements}
The image in Fig.\ref{fgr:fig1} was recorded inside a MBE vessel in collaboration with P. Atkinson, M. Eddrief, V. Etgens, F. Finocchi from Institut des NanoSciences de Paris and H. Khemliche, A. Momeni A.G. Borisov, A. Zugarramurdi from ISMO. We are grateful to Hynd Remita from the Institut de Chimie Physique for the irradiation of the LiF samples and Andrew Mayne for his help in reading the drafts.
This work has been funded by the Agence Nationale de la Recherche (contract ANR-2011-EMMA-003-01) and Chinese Scholarship Council (CSC) Grant reference number 201806180025.

%%%END OF MAIN TEXT%%%

%The \balance command can be used to balance the columns on the final page if desired. It should be placed anywhere within the first column of the last page.

\balance

%If notes are included in your references you can change the title from 'References' to 'Notes and references' using the following command:
%\renewcommand\refname{Notes and references}

%%%REFERENCES%%%
\bibliography{rsc} %You need to replace "rsc" on this line with the name of your .bib file
\bibliographystyle{unsrt.bst}
\end{document}